\documentclass[fleqn,usenatbib]{mnras}

\usepackage[normalem]{ulem}

\usepackage[T1]{fontenc}

\DeclareRobustCommand{\VAN}[3]{#2}
\let\VANthebibliography\thebibliography
\def\thebibliography{\DeclareRobustCommand{\VAN}[3]{##3}\VANthebibliography}

\usepackage{graphicx}	
\usepackage{amsmath}	
\usepackage{amssymb}	
\usepackage{float}
\usepackage{placeins}

\title[The build up of the quenched galaxy population since cosmic noon]{The role of mass and environment in the build up of the quenched galaxy population since cosmic noon}

\author[E. Taylor et al.]{Elizabeth Taylor,$^{1}$ Omar Almaini,$^{1}$  Michael Merrifield,$^{1}$
David Maltby,$^{1}$ Vivienne Wild,$^{2}$ \newauthor William G. Hartley, $^{3}$ Kate Rowlands$^{4}$
\\
$^{1}$University of Nottingham, School of Physics and Astronomy, Nottingham, NG7 2RD, U.K\\
$^{2}$School of Physics and Astronomy, University of St Andrews, North Haugh, St Andrews, KY16 9SS, U.K\\
$^{3}$Department of Astronomy, University of Geneva, ch. d’Ecogia 16, CH-1290 Versoix, Switzerland\\
$^{4}$Johns Hopkins University, Department of Physics and Astronomy, Baltimore, MD 21218, USA
}

\date{Accepted XXX. Received YYY; in original form ZZZ}

\pubyear{2022}

\begin{document}
\label{firstpage}
\pagerange{\pageref{firstpage}--\pageref{lastpage}}
\maketitle

\begin{abstract}
We conduct the first study of how the relative quenching probability of galaxies depends on environment over the redshift range $0.5 < z < 3$, using data from the UKIDSS Ultra-Deep Survey. By constructing the stellar mass functions for quiescent and post-starburst (PSB) galaxies in high, medium and low density environments to $z = 3$, we find an excess of quenched galaxies in dense environments out to at least $z \sim 2$. Using the growth rate in the number of quenched galaxies, combined with the star-forming galaxy mass function, we calculate the probability that a given star-forming galaxy is quenched per unit time. We find a significantly higher quenching rate in dense environments (at a given stellar mass) at all redshifts. Massive galaxies (M$_* > 10^{10.7}$ M$_{\odot}$) are on average 1.7 $\pm$ 0.2 times more likely to quench per Gyr in the densest third of environments compared to the sparsest third. Finally, we compare the quiescent galaxy growth rate to the rate at which galaxies pass through a PSB phase. Assuming a visibility timescale of 500 Myr, we find that the PSB route can explain $\sim$ 50\%  of
the growth in the quiescent population at high stellar mass (M$_* > 10^{10.7}$ M$_{\odot}$) in the redshift range $0.5 < z < 3$,  and potentially all of the growth at lower stellar masses.
\end{abstract}




\section{Introduction} \label{section:intro}
One of the areas of great interest in astrophysics is how galaxies transition from being blue, highly star-forming structures to red, passive systems. Recent surveys have enabled measurements of both star-forming and quiescent galaxies out to high redshifts, allowing us to determine how the fraction of these systems evolves with time \citep[e.g.][]{bell_nearly_2004, peng_mass_2010, muzzin_evolution_2013, papovich_effects_2018, leja_new_2020}. Once a galaxy lies on the red sequence, it is likely to remain there, leading to an overall build up of the quiescent population.

There is still a gap in our understanding, however, as to what leads to the cessation of star formation in blue galaxies, causing their evolution onto the red sequence. This process is commonly referred to as quenching, and comprises many possible physical mechanisms that could theoretically lead to a gas-rich galaxy experiencing a drop in star formation. These mechanisms include external processes such as the removal of gas from galaxies via strangulation \citep[e.g.][]{larson_evolution_1980, bosch_importance_2008} or ram pressure stripping \citep[]{gunn_infall_1972}, and internal processes such as AGN feedback \citep[e.g.][]{silk_quasars_1998, hopkins_unified_2006}.

One way to investigate quenching is to study the evolution of the galaxy stellar mass functions (SMFs) that define the number density of galaxies of different masses. SMFs allow us to trace the assembly of stellar mass through cosmic time, and thus they provide a useful tool for studying quenching by quantifying the transformation of star-forming galaxies into quiescent systems \citep[e.g.][]{bell_optical_2003, baldry_galaxy_2008, pozzetti_zcosmos_2010, moustakas_primus_2013}. Star formation in massive galaxies was at its peak from $z \sim 3$ to $z \sim 1$, a period often referred to as "cosmic noon", and over this relatively short period of time these systems formed roughly half their current stellar mass \citep[see][]{schreiber_star-forming_2020}. Furthermore, there is strong evidence that most massive quiescent galaxies were already in place by $z \sim 1$ \citep[e.g.][]{ilbert_mass_2013, muzzin_evolution_2013, wright_gamag10-cosmos3d-hst_2018, mcleod_evolution_2021}. \citet{ilbert_mass_2013} found that the number density of quiescent galaxies increases from $z \sim 3$ to $z \sim 1$ over all stellar masses, with no significant evolution at the high mass end below $z \sim 1$. Similarly, \citet{mcleod_evolution_2021} found that while passive galaxies contributed $<$ 10 per cent to the total stellar mass density of the Universe at $z \sim 3$, they dominate by $z \sim 0.75$.

Quenching is often broadly grouped into mass quenching and environment quenching \citep[e.g][]{peng_mass_2010, muzzin_gemini_2012, papovich_effects_2018}. \citet{peng_mass_2010} found that the effects of mass and environment on quenching were completely separable out to $z \sim 1$, and were able to reproduce the evolution of the SMFs of blue and red galaxy populations with this quenching formalism. They found that the efficiency of mass quenching is proportional to the star formation rate (SFR), and that the efficiency of environment quenching depends on the galaxy density field, with galaxies in the highest density environments having a higher likelihood of being quenched \citep[see also][]{kovac_zcosmos_2014, knobel_quenching_2015}. In contrast, some studies of groups and clusters find that the same formalism used by Peng et al. may not hold past $z \sim 0.7$ \citep[e.g.][]{balogh_evidence_2016, mcnab_gogreen_2021}. Studies at higher redshifts suggest either a declining influence from environmental quenching \citep[e.g.][]{nantais_stellar_2016} or potentially no additional quenching due to environment \citep[]{darvish_effects_2016}.
Furthermore, \citet{de_lucia_environmental_2012} propose that in the real Universe it is unexpected that mass and environment separability holds at any redshift, as the two are physically connected \citep[see also][]{pintos-castro_evolution_2019}. Observationally, however, the role played by environment in quenching galaxies remains unclear at $z > 1$, which provides motivation for the current study.

Studying galaxies in transitionary stages may provide the key to understanding the processes responsible for terminating star formation. One such class are post-starburst galaxies \citep[PSBs, e.g.][]{dressler_spectroscopy_1983, tran_field_2004, wild_post-starburst_2009, french_evolution_2021}, which are thought to be systems that have undergone a recent, major burst of star formation (or an extended period of high star formation) followed by rapid quenching. PSBs are typically identified by their spectra showing both strong Balmer absorption lines typical of young A/F stars, and weak or absent emission lines. \cite{wild_evolution_2016} found that they are rare at all epochs: less than 1\% of local massive galaxies are PSBs, with this number rising to $\sim$ 5\% at $z \sim 2$. Studies have found that PSBs could potentially evolve to produce up to $\sim$ 50 per cent of the massive red sequence, depending on the assumptions made about their visibility timescales and star formation histories \citep[e.g][]{wild_evolution_2016, belli_mosfire_2019, wild_star_2020}. Various authors have found that PSBs are more abundant in denser environments to $z \sim 1$, suggesting a strong link to environmental quenching \citep[e.g][]{vergani_kgalaxies_2010, muzzin_gemini_2012, socolovsky_enhancement_2018, paccagnella_strong_2019}. To date, however, there have been no studies of the link between PSBs and environment at higher redshifts ($z > 1.5$), when massive (M$_* > 10^{10}$ M$_{\odot}$) PSBs were much more common \citep[]{wild_evolution_2016}. Studies of this nature could provide insight into the mechanisms responsible for quenching massive galaxies at this crucial epoch.

In this paper, we investigate the influence  of stellar mass and environment on the build up of the quiescent population since $z = 3$. We revisit the evolution of PSBs and the galaxy mass functions, but now (for the first time) separated by environment. We then calculate both the growth rate of the quiescent population and the quenching rate in separate redshift, mass and environment bins to analyse whether these factors are indeed separable out to high redshift. The deep imaging from the Ultra-Deep Survey (UDS) is ideal for this study, allowing us to probe typical galaxies ($\sim$ 10$^{10}$ M$_\odot$) to $z \sim 3$, with sufficient volume to enable samples to be split by stellar mass, redshift and environment.

The structure of this paper is as follows: In Section \ref{section:data} we present our data, the galaxy classification  method, and the environment measurements. In Section \ref{section:results} we present our results on the quenching rates in different environments, before performing a series of robustness tests in Section \ref{section:rob}. In Section \ref{section:disc} we explore the contribution of PSBs to the build-up of the passive galaxy population. We end with our conclusions and a brief summary in Section \ref{section:conc}. Throughout this paper, we adopt the AB magnitude system and a flat $\Lambda$CDM cosmology with $\Omega_M$ = 0.3, $\Omega_{\Lambda}$ = 0.7, and $H_0 = 100$ \textit{h} kms$^{-1}$ Mpc$^{-1}$ where $h$ = 0.7.

\begin{figure}
	\includegraphics[width=\columnwidth]{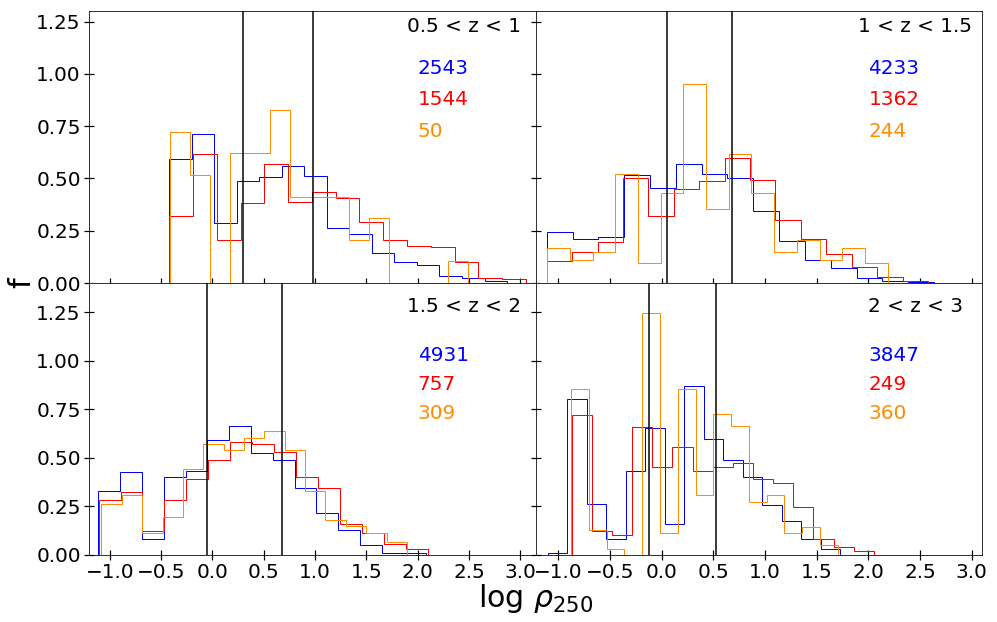}
    \caption{Histograms of environmental density $\rho_{250}$ in our four redshift intervals, displayed separately for the star-forming (blue), quiescent (red) and post-starburst (yellow) galaxy populations. Our $\rho_{250}$ values are calculated as described in Section \ref{section:env}. Solid black vertical lines in each panel represent the $\rho_{250}$ values at which we split the total (combined) population to provide three density bins containing equal  numbers of galaxies with M$_* > 10^{9.91}$ M$_{\odot}$. The total number of galaxies in each population is indicated. }
    \label{fig:hist}
\end{figure}

\begin{figure*}
	\includegraphics[width=\textwidth]{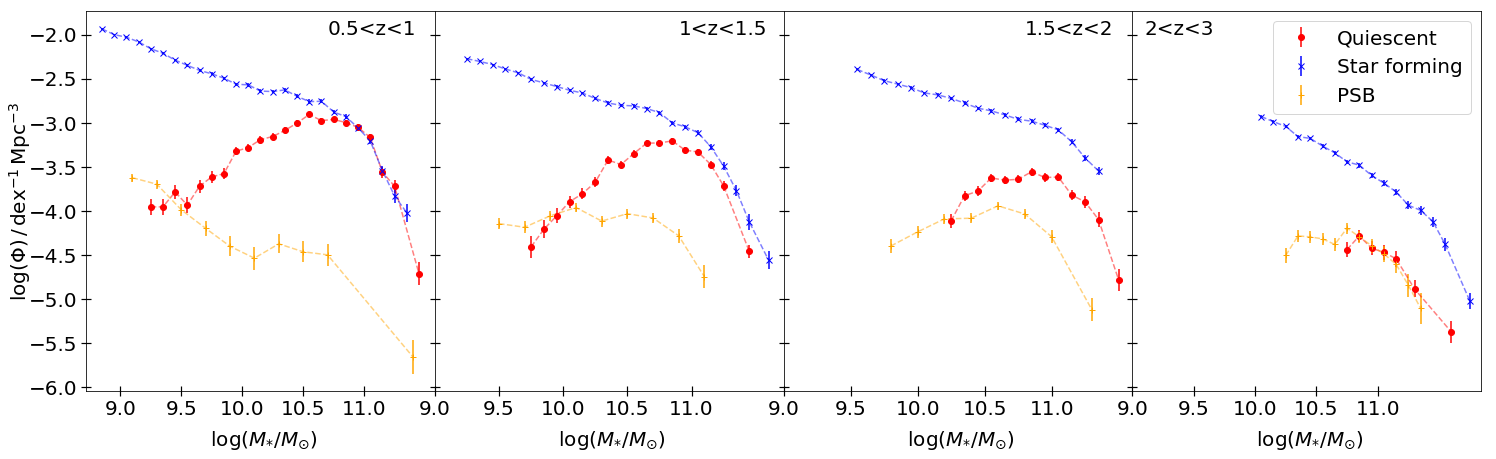}
    \caption{Global stellar mass functions for the quiescent (red), star-forming (blue) and PSB (yellow) populations as a function of redshift. The mass functions are cut off at the 90\% mass completeness limit for each population, evaluated at the upper end of each redshift bin. Poisson error bars are shown.}
    \label{fig:glob}
\end{figure*}

\section{Data and Method} \label{section:data}

\subsection{The UDS} \label{section:UDS}

The UDS is the deepest NIR survey conducted over $\sim$ 0.8 deg$^2$, and the deepest component of the United Kingdom Infrared Telescope (UKIRT) Infrared Deep Sky Survey \citep[UKIDSS;][]{lawrence_ukirt_2007}. We use the UDS Data Release 11 (DR11) catalogue, which reaches a 5$\sigma$ limiting depth of \emph{J} = 25.6, \emph{H} = 25.1, and \emph{K} = 25.3 in 2-arcsec diameter apertures. Further details can be found in \citet{almaini_massive_2017} and \citet{wilkinson_starburst_2021}. Full details of the DR11 catalogue and photometric redshifts will be presented in Almaini et al. (in prep). 

To complement the UKIDSS imaging, we use additional deep photometric data in 9 other bands: \emph{B}, \emph{V}, \emph{R}, \emph{i'} and \emph{z'}-band optical observations from the Subaru XMM-Newton Deep Survey \citep[SXDS,][]{furusawa_subaru_2008}, \emph{U'}-band photometry from the CFHT Megacam, mid-infrared photometry (3.6$\mu$m and 4.5$\mu$m) from the \emph{Spitzer} UDS Legacy Program (SpUDS, PI:Dunlop) and deep \emph{Y}-band data from the VISTA VIDEO survey \citep[]{jarvis_vista_2013}. The area of the UDS covered by all 12 bands is 0.62 deg$^2$.

We use photometric redshifts computed using the method outlined in \citet{simpson_prevalence_2013}, using the EAzY code \citep[]{brammer_eazy_2008} to fit a linear combination of template SEDs to our photometry at each redshift grid point. Our set-up broadly follows the default EAzY configuration for the 12 Flexible Stellar Population Synthesis (FSPS) SED components \citep[]{conroy_propagation_2010}, with the addition of three simple stellar population (SSP) templates. This combination gave the best results in terms of the scatter and outlier fraction when compared to the available spectroscopic redshifts. The three SSP models have ages of 20, 50 and 150 Myr, using a Chabrier IMF and sub-solar metallicity (0.2 solar), and represent recent bursts of star formation to complement the complex and continuous star-formation histories of the FSPS templates. We found that varying the metallicities of these bursts was unimportant to the results. The UDS contains $\sim$ 8000 sources with secure spectroscopic redshifts, and comparison of the photometric redshifts yields a mean absolute dispersion in (z$_{phot}$ - z$_{spec}$)/(1 + z) of $\sigma_{NMAD}$ = 0.019 with an outlier fraction (|$\Delta$z|/(1 + z) > 0.15) of $\sim$ 3$\%$. Further detail on the determination of photometric redshifts can be found in \cite{simpson_prevalence_2013} and Hartley et al. (in prep.).

Stellar masses are calculated using a Bayesian analysis, following the method outlined in \citet{wild_evolution_2016}. 10's of thousands of \citet{bruzual_stellar_2003} population synthesis models are constructed with a range of star-formation histories, and then fitted to our supercolours (see Section \ref{section:psbselection}). The resulting internal errors in the stellar masses are typically $\pm$ 0.1 dex at fixed redshift, allowing for the degeneracy between fitted parameters \citep[see][]{wild_evolution_2016}. Further discussion of systematic stellar mass uncertainties can be found within \citet{almaini_massive_2017}. The 90\% stellar mass completeness limits are calculated using the method of \cite{pozzetti_zcosmos_2010} and are given for the different galaxy populations in Table \ref{tab:1}. We note that these limits are conservative, as they are evaluated at the upper end of each redshift interval, as this allows the samples to be volume complete. Star formation rates (SFR) are generated by comparison of the observed galaxy supercolours (discussed below, see section \ref{section:psbselection}) with those of the Bruzual \& Charlot models \citep[]{wild_evolution_2016, wilkinson_starburst_2021}.


\begin{table}
\centering
	\caption{90\% stellar mass completeness limits in units log(M$_*$/M$_{\odot}$) for the star-forming, quiescent and PSB populations in each redshift bin. Mass completeness limits are calculated following the method of \citet{pozzetti_zcosmos_2010}.}
	\label{tab:1}
	\scalebox{0.9}{\hskip-0.5cm	\begin{tabular}{|c||c|c|c|c|} 
		\hline
		& 0.5 $<$ z $<$ 1 & 1 $<$ z $<$ 1.5 & 1.5 $<$ z $<$ 2 & 2 $<$ z $<$ 3 \\
    \hline
    \hline
    Star-forming & 8.85 & 9.22 & 9.53 & 9.95 \\
    \hline
    Quiescent & 9.2 & 9.74 & 10.17 & 10.72 \\
    \hline
    PSB & 9.06 & 9.48 & 9.81 & 10.22 \\
    \hline
\end{tabular}}

\end{table}

\subsection{Galaxy Classification} \label{section:psbselection}
To separate star-forming galaxies, passive galaxies, and PSBs, we use the principal component analysis (PCA) technique established by \citet{wild_new_2014, wild_evolution_2016} \citep[with further spectroscopic confirmation by][]{maltby_identification_2016}. In brief, the aim of the PCA method is to describe the broad range of galaxy SEDs using the linear combination of a small number of components. It is found that three components are needed to sufficiently account for the variance in SEDs, with the amplitude of each component being termed a `supercolour' (SC). The first two supercolours, SC1 and SC2, are the most useful in determining whether a galaxy is a good PSB candidate. SC1 correlates with sSFR and mean stellar age, and SC2 correlates with the fraction of a galaxy's stellar mass formed within the last billion years. Galaxies can be classified based on their position in a SC1-SC2 diagram, where the population boundaries are determined by comparison to model SEDs and spectroscopy. PSBs are identified as galaxies with a low value of SC1 (low sSFR) and a high value of SC2 ($>$ 10\% stellar mass built up in the last Gyr). The “dusty” star-forming class identified by \cite{wild_new_2014} are combined with other star-forming galaxies for our analysis. For a more detailed explanation of the supercolour analysis, see \citet{wild_new_2014}. 

This technique was originally established using the UDS DR8 catalogue, but here we use an updated SC analysis applied to the full DR11 catalogue, explained in more detail in \citet{wilkinson_starburst_2021}. The main modification made for the DR11 catalogue is to extend the  redshift range (adding in analysis for $2 < z < 3$), which required a slight modification to the  rest-frame wavelength range analysed. The DR11 catalogue is also deeper, with more accurate photometric redshifts (see Section \ref{section:UDS}).

\subsection{Environment Measurements} \label{section:env}

We measure the galaxy environments in a statistical manner, following a similar  method to \cite{lani_evidence_2013}, but applied to the newer DR11 catalogue and photometric redshifts. In brief, each galaxy has their local density ($\rho_{250}$) measured by construction of a cylinder centred on the galaxy of fixed radius 250 kpc and depth equivalent to $\pm$ 0.5 Gyr along the redshift direction. The number of galaxies within the cylinder is counted and normalised to the expected number, given the density of galaxies in the wider field within the same redshift range (allowing for holes and edges). The depth of $\pm$ 0.5 Gyr is assigned as it is significantly larger than the 1$\sigma$ uncertainty of $\delta$z = 0.0187(1+z) on the photometric redshifts (Almaini et al., in preparation): for example, at $z = 1$, the 1$\sigma$ redshift error corresponds to 0.151 Gyr, and hence a depth of 0.5 Gyrs corresponds to 3.3$\sigma$. It should be noted we cannot say with certainty if an individual galaxy is in a cluster or in the field; however, due to the large sample size, we can  group galaxies into high- and low-density environments, to investigate density-dependent behaviour in a statistical manner. The robustness of our environmental measurements is explored further in Section \ref{section:rob}. 

We split the catalogue into broad redshift intervals, and within each redshift interval galaxies are sorted into three environmental bins: high, medium, and low density. The three categories are obtained by sorting galaxies by $\rho_{250}$ and splitting into equal thirds (equal numbers of galaxies in each category) for galaxies with stellar mass M$_* > 10^{9.91}$ M$_{\odot}$, as shown in Figure \ref{fig:hist}. This stellar mass represents the 90 per cent completeness limit for galaxies at our upper redshift bound ($z = 3$), as described in Section \ref{section:UDS}. We measure environments using only galaxies above this mass limit, to ensure that environments are compared with reference to a similar galaxy population at all redshifts. Galaxies below this mass limit are also assigned environmental bins, but using environments defined by the galaxies with M$_* > 10^{9.91}$ M$_{\odot}$. In the later analysis, we are implicitly assuming, for simplicity, that galaxies will remain within their broad density thirds as they evolve, even if the galaxy itself is transformed or quenched. 
The impact of these assumptions will be explored and tested further in Section \ref{section:rob}.


\begin{figure*}
	\includegraphics[width=\textwidth]{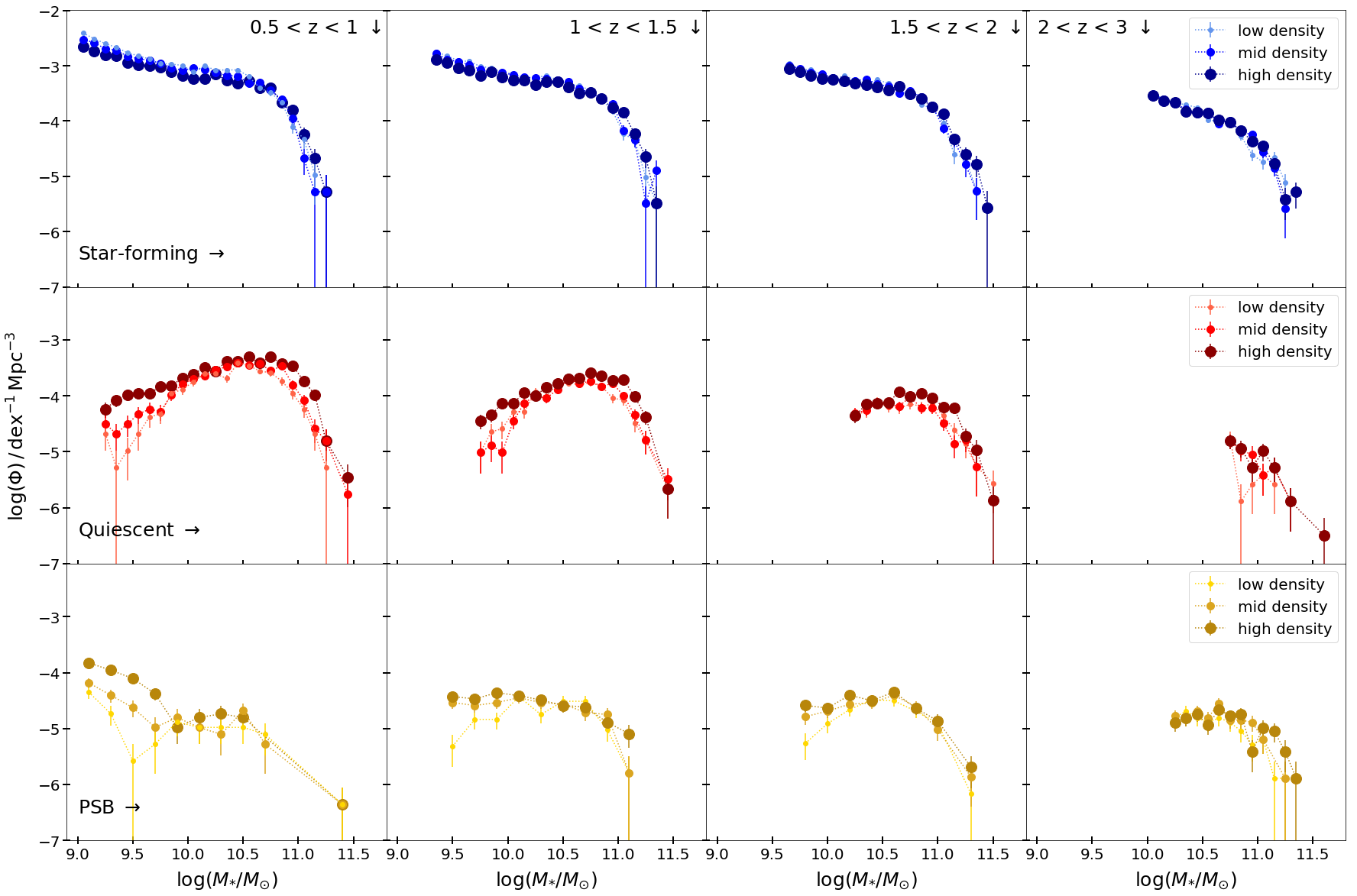}
    \caption{Global stellar mass functions for the star-forming (blue, row 1), quiescent (red, row 2) and PSB (yellow, row 3) populations, shown in four redshift intervals, each split into three environmental bins. The three environmental bins are obtained using the combined galaxy population, sorted by $\rho_{250}$ (see Section \ref{section:env}). Poisson error bars are shown. The mass functions are cut off at the 90\% mass completeness limit for each population for the given redshift bin.}
    \label{fig:envmfs}
\end{figure*}

\section{Results} \label{section:results}

\subsection{Global mass functions} \label{section:gmf}

In Figure \ref{fig:glob}, we show the global SMFs of star-forming, quiescent and PSB galaxies as a function of redshift. The number density uncertainties are simply the Poisson counting errors. We see the expected result \citep[see e.g.][]{muzzin_evolution_2013}: the star-forming mass function shape remains almost constant across cosmic time, while the quiescent mass function shows a large build up in number density towards low redshift, especially at the high-mass end. At redshifts below $z = 1$, the quiescent galaxy SMF shows a  more significant build up at the low mass end. We see a change in the PSB mass function across redshift, confirming the results of  \citet{wild_evolution_2016}; at high redshift ($z > 1.5$), the PSB mass function is similar to that of the quiescent population, but at lower redshifts there is an upturn in the relative number of PSBs at lower stellar masses, more closely resembling the shape of the star-forming galaxy mass function  \citep[see also][]{maltby_structure_2018}. Our extension to $z = 3$ confirms that the PSB and quiescent mass functions are essentially the same at the highest redshifts; performing a Kolmogorov–Smirnov (KS) test on the shape of the PSB and quiescent SMFs in the redshift range $2 < z < 3$, we cannot reject the null hypothesis that they are drawn from the same underlying distribution (p = 0.27). This similarity suggests that quiescent galaxies in this redshift bin could be quenched by the same mechanisms that quenched PSBs. The upturn in the PSB mass function at low redshifts suggests there may be two separate pathways to form PSBs, with a mechanism that creates low-mass PSBs acting preferentially at $z < 1$ \citep[see][]{wild_evolution_2016, maltby_structure_2018}.

\begin{figure*}
	\includegraphics[width=\textwidth]{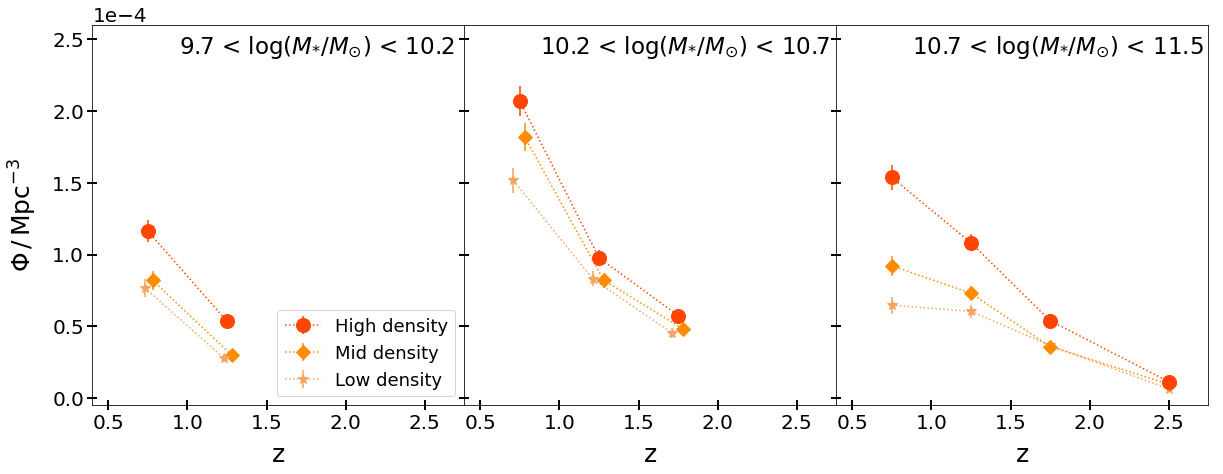}
    \caption{The evolution of passive galaxy number density as a function of redshift, split by mass and environment. Density bins are given in the legend, with darker colours and larger markers representing denser environments. Points are plotted at the midpoint of their respective redshift interval and are offset slightly for clarity. Poisson error bars are shown.}
    \label{fig:nde}
\end{figure*}

\subsection{Mass functions split by environment} \label{section:emf}

Figure \ref{fig:envmfs}, row 1 shows the SMFs for the star-forming galaxy populations as a function of redshift and environment. The SMFs are very similar overall in amplitude as expected, however there is evidence for a relative deficit of low mass star-forming galaxies in high-density environments at low redshift ($z < 1$). Comparing star-forming galaxies in high-density and low-density environments, at low stellar mass (M$_* < 10^{10}$ M$_{\odot}$) and low redshift ($z < 1$) the difference is formally significant to 16$\sigma$.

Figure \ref{fig:envmfs}, row 2 shows the SMFs for the quiescent galaxies as a function of redshift and environment. As expected, we see growth in the number density of quiescent galaxies towards low redshift. In the redshift interval $1.5 < z < 3$ the mass functions across all three environments are similar. We see that from $z = 2.0$ to $z = 0.5$ there is evidence for an increase in the relative abundance of quiescent galaxies in dense environments versus sparse environments, suggesting that  quenching is more effective in dense environments towards lower redshift. This is especially significant ($\sim$ 7$\sigma$) for low mass quiescent galaxies (M$_* < 10^{10}$ M$_{\odot}$) at $z < 1.5$. Similar results have been found in earlier work \citep[e.g.][]{mortlock_deconstructing_2015, annunziatella_clash-vlt_2016, van_der_burg_stellar_2018, socolovsky_enhancement_2018, van_der_burg_gogreen_2020}.

Figure \ref{fig:envmfs}, row 3 shows the SMFs for PSBs within the UDS, as a function of redshift and environment. There is only marginal evidence for an excess of PSBs in the dense environments at high stellar mass, but at low stellar mass (M$_* < 10^{10}$ M$_{\odot}$) we see an up-turn in the relative abundance of PSBs in dense environments at $z < 1.5$, mirroring a similar trend in the older quiescent galaxies. In the redshift bin $0.5 < z < 1$, for low-mass galaxies (M$_* < 10^{10}$ M$_{\odot}$), the excess of PSBs in high versus low-density environments is formally significant to $\sim$ 9$\sigma$. The PSB SMFs are similar in all environments at $z > 2$, indicating that environment may only play a role in rapid quenching at lower redshifts. These results are consistent with the work of \cite{socolovsky_enhancement_2018}, who also found an excess of lower mass PSBs in dense environments at $z < 1$, in their case by splitting the UDS galaxies into `cluster' and `field' environments using a friends-of-friends algorithm.

\subsection{The growth in the number of passive galaxies and the dependence on environment} \label{section:gr}

\begin{figure*}
	\includegraphics[width=\textwidth]{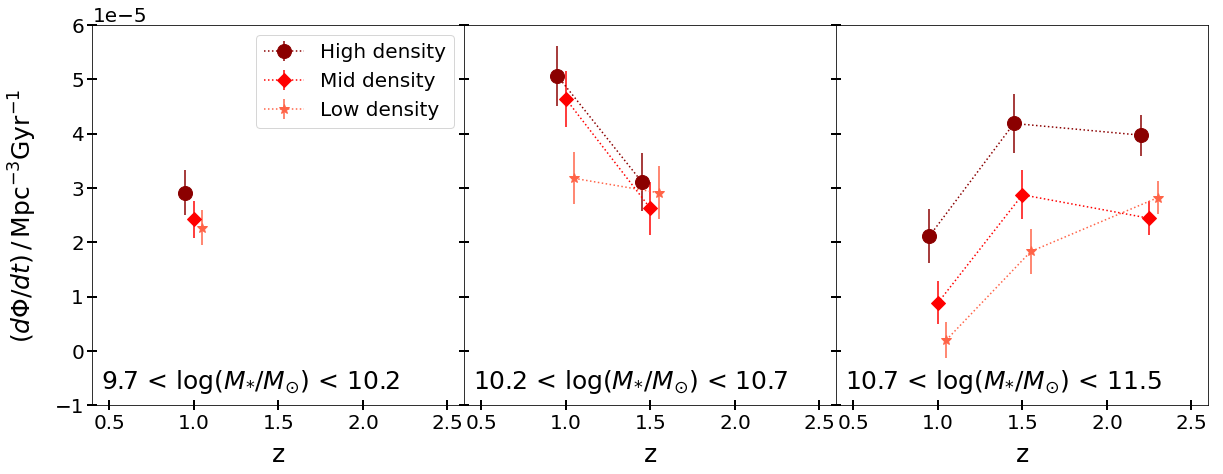}
    \caption{The number density growth rate of passive galaxies as a function of redshift, split by mass and environment. Density bins are given in the legend, with darker colours and larger markers representing denser environments. Points are plotted at the boundary between the two redshift intervals used in their calculation and are offset slightly for clarity. Errors are propagated from the Poisson error.}
    \label{fig:gr}
\end{figure*}

\begin{figure*}
	\includegraphics[width=\textwidth]{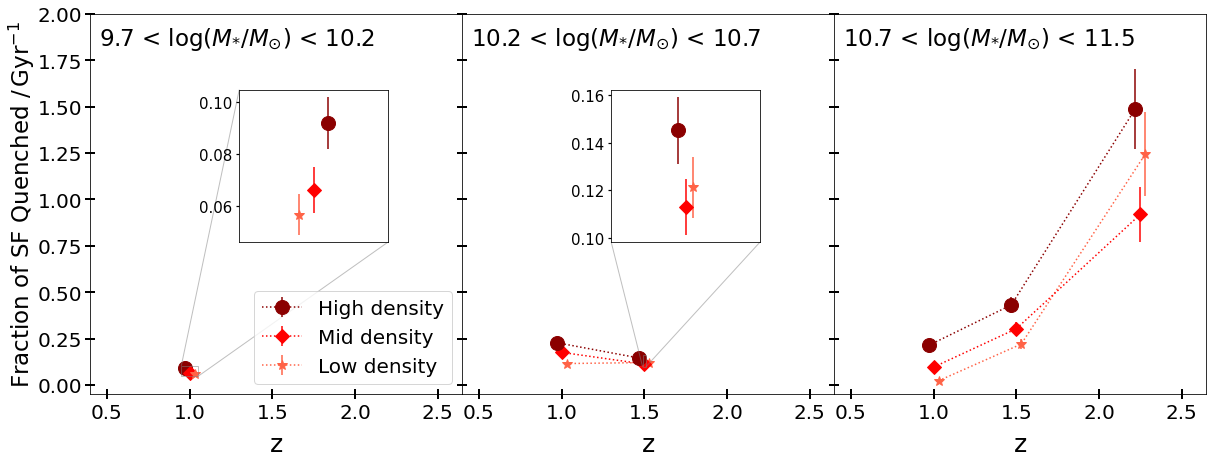}
    \caption{The fraction of star-forming galaxies quenched per Gyr as a function of redshift, shown for three density bins and over three ranges in stellar mass. Points are plotted at the boundary between the two redshift intervals used in their calculation and are offset slightly for clarity. Errors are propagated from the Poisson error.}
    \label{fig:qf}
\end{figure*}

In this Section and Section \ref{section:qf} we investigate the rate of growth in the number density of passive galaxies in order to estimate a quenching rate. Given the relatively small numbers of PSBs, we combine PSBs and older quiescent  galaxies into a single `passive' galaxy population.
Figure \ref{fig:nde} shows the resulting evolution in number density of the combined passive population from $z = 3$ to $z = 0.5$. The three panels compare the number density evolution split by environment (as defined in Section \ref{section:env}) in three mass bins (increasing from left to right). We see that the number density of quenched galaxies is higher in high-density environments at all redshifts, especially at the highest masses. The slope of the number density evolution (quantified in Figure \ref{fig:gr}) is also steepest for the highest density environments, indicating star-forming galaxies are more likely to be quenched over a given timescale. 

Figure \ref{fig:gr} shows the growth rate in the passive galaxy number density within the UDS. We calculate this rate by approximating the derivative of the number density evolution:

\begin{equation} \label{eq:gr}
      \frac{d\phi_{\textrm{red}}}{dt} \approx \frac{\phi_{\textrm{red, low z}}\, -\, \phi_{\textrm{red, high z}}}{\Delta t} ,
\end{equation}

where $\phi_{\textrm{red, low z}}$ and $\phi_{\textrm{red, high z}}$ are the passive galaxy number densities in the lower and upper redshift bins respectively, and $\Delta t$ is the difference in Gyrs between the midpoint of each redshift bin. Errors are propagated from the Poisson counting errors. There is evidence that high-density environments show the highest growth rate at all stellar masses. This trend is most significant in the highest mass bin: comparing low and high-density environments, the differences are significant at 3.3, 3.6 and 2.9$\sigma$ for $z = 1$, $z = 1.5$ and $z = 2$ respectively.
The growth rate in the number of  high-mass passive galaxies (M$_* > 10^{10.7}$ M$_{\odot}$) in the densest environments shows only a small decrease across cosmic time, which seems to imply that the mechanisms causing enhanced quenching for this population are in place from at least $z = 3$.

\subsection{The probability of quenching and its dependence on environment} \label{section:qf}

In Section \ref{section:gr} we calculated the growth rate in the passive galaxy mass function, but in order to physically interpret this rate as quenching probability we need to consider how many star-forming galaxies are available to be quenched. Figure \ref{fig:qf} shows the fraction of star-forming galaxies in a given mass and environment bin that are quenched per Gyr, given by:

\begin{equation} \label{eq:qf}
    \lambda = \frac{d\phi_{\textrm{red}}}{dt}\frac{1}{\phi_{\textrm{SF, high z}}} = \frac{\phi_{\textrm{red, low z}}\, -\, \phi_{\textrm{red, high z}}}{\phi_{\textrm{SF, high z}}\, \Delta t} ,
\end{equation}

Equation \ref{eq:qf} is the growth rate (Equation \ref{eq:gr}) divided by the number density of star-forming galaxies in the higher redshift bin -- in other words, we calculate the fraction of star-forming galaxies that were available to be quenched from the last timestep that went on to be quenched by the next redshift interval. This simple methodology neglects merging, and also ignores star-forming galaxies that are added to a given mass bin and quenched between the intervening time intervals; the potential impact of these assumptions is explored further in Section \ref{section:rob}.

At all redshifts and stellar masses we find the quenching rate is higher in the densest environments (see Table \ref{tab:2}). There is also evidence that the quenched fraction increases with stellar mass, as we see an upward trend at a given redshift from the left to right panels in Figure \ref{fig:qf}.

Perhaps the most striking trend in Figure \ref{fig:qf} is the enhanced quenching probability in dense environments at a given redshift and stellar mass.
The final row in Table \ref{tab:2} gives the ratio of the quenching rate in high-density compared  to low-density environments.
On average, the highest mass galaxies (M$_* > 10^{10.7}$ M$_{\odot}$) are 1.7 $\pm$ 0.2 times more likely to quench per Gyr in the densest third of environments than in the lowest density third. This value is a weighted mean of the high- and low-density ratios in the three redshift bins. If we use all mass bins we find that a galaxy is 1.5 $\pm$ 0.1 more likely to quench in the densest third of environments compared to the low-density third. 

\subsection{Comparison to low redshift empirical trends} \label{section:lit}

\begin{figure*}
    \centering
    \includegraphics[width=\textwidth]{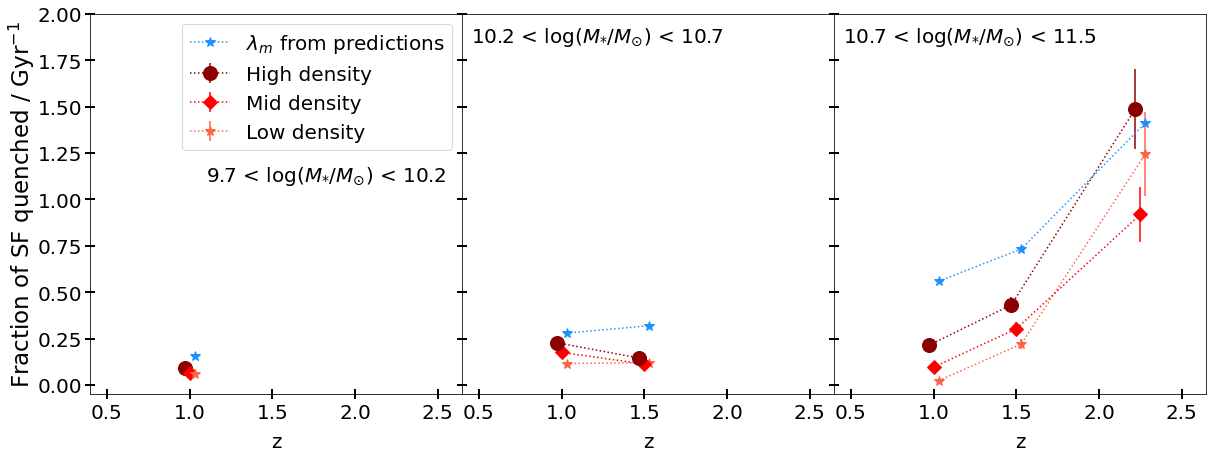}
    \caption{A simple comparison of the quenched fraction of star-forming galaxies (red points) to $\lambda_m$ predictions using Equation \ref{eq:pred} (blue points). Points are plotted at the midpoint of their respective redshift interval and are offset slightly for clarity. Errors are propagated from the Poisson error.}
    \label{fig:lm}
\end{figure*}

One area of great interest in the literature is  whether the effects of galaxy environment and stellar mass on quenching are separable, as this may shed light on the likely physical mechanisms resposible. \citet{peng_mass_2010} looked at this in detail, and concluded that mass quenching and environment quenching are indeed distinct out to $z \sim 1$. Peng et al. identify a mass quenching rate $\lambda_m$ (their equation 21) which they find depends only on the SFR as follows

\begin{equation} \label{eq:pred}
    \lambda_m = \left( \frac{\textrm{SFR}}{40\, M_{\odot}\, \textrm{yr}^{-1}} \right) \textrm{Gyr}^{-1} .
\end{equation}

To compare our observed overall quenching rate $\lambda$ with the predicted $\lambda_m$ from Equation \ref{eq:pred}, we use the median SFR (see Section \ref{section:data}) for star-forming galaxies in low-density environments in each of the mass and redshift bins, divided by 40 M$_{\odot}$ yr $^{-1}$ to produce a prediction the mass quenching rate.
 We use galaxies in the low-density environments only,  to minimise the potential influence of environment. This simple comparison is plotted in Figure \ref{fig:lm}, where the observed and predicted quenching rates are shown in red and blue respectively. In general terms we find that our observed quenching rates broadly follow the mass-quenching predictions from \citet{peng_mass_2010}, steadily
increasing with both stellar mass and redshift.  Formally,  the predictions slightly over-predict the observed quenching rates at low densities, particularly at low stellar mass. As noted in Section \ref{section:intro}, other studies have found evidence for a more complex relationship between quenching and environment at $z > 1$ \citep[e.g][]{balogh_evidence_2016, kawinwanichakij_effect_2017, mao_revealing_2022}, but a detailed comparison is beyond the scope of this work. In broad terms, however, we find that the evolution of the quenching rates do broadly follow the trends expected if quenching is driven by two independent mechanisms: one related to SFR, and the other related to environment.


\section{Robustness tests} \label{section:rob}

\begin{table*}
	\centering
	\caption{Fraction of star-forming galaxies quenched per Gyr for each mass and density bin, at each redshift (plotted in Figure \ref{fig:qf}). The final row shows the ratio of high-density to low-density points for each bin.}
	\label{tab:2}
	\scalebox{0.94}{\hskip-0.5cm
	\begin{tabular}{ | c | c | c | c | c | c | c |  } 
		\hline
		 &  9.7 $<$ log(M$_*$/M$_{\odot}$) $<$ 10.2 & \multicolumn{2}{c}{10.2 $<$ log(M$_*$/M$_{\odot}$) $<$ 10.7} & \multicolumn{3}{c}{10.7 $<$ log(M$_*$/M$_{\odot}$) $<$ 11.5}\\
    \hline
    & z = 1 & z = 1 & z = 1.5 & z = 1 & z = 1.5 & z = 2.25  \\
    \hline
    \hline
    Low-density & 0.057 $\pm$ 0.008 &  0.116 $\pm$ 0.012 & 0.121 $\pm$ 0.013 & 0.023 $\pm$ 0.003 & 0.220 $\pm$ 0.028 & 1.244 $\pm$ 0.227 \\
    \hline
    Medium-density & 0.066 $\pm$ 0.009 & 0.176 $\pm$ 0.016 & 0.113 $\pm$ 0.012 & 0.097 $\pm$ 0.011 & 0.303 $\pm$ 0.037 & 0.919 $\pm$ 0.148  \\
    \hline
    High-density & 0.092 $\pm$ 0.010 & 0.228$\pm$ 0.020 & 0.145 $\pm$ 0.014 & 0.213 $\pm$ 0.021 & 0.432 $\pm$ 0.045 & 1.488 $\pm$ 0.217  \\
    \hline
    High-density / Low-density & 1.620 $\pm$ 0.285 & 1.972 $\pm$ 0.249 & 1.198 $\pm$ 0.171 & 9.279 $\pm$ 1.522 & 1.957 $\pm$ 0.322 & 1.196 $\pm$ 0.279  \\ 
    \hline
    
\end{tabular}}
\end{table*}

In this section we explore some of the simplifying assumptions that we have made to this point, and discuss some of the resulting caveats.

\subsection{Growth through mergers}\label{section:merge}
One issue that we have not considered thus far is that galaxies no longer forming stars may still increase in stellar mass via mergers, and consequently move to higher mass bins, contributing to a change in the mass functions. To estimate the magnitude of the effect mergers may have on the derived quenching rates, we use the fractional (major) merger rate found for the UDS by \citet{mundy_consistent_2017}, which at these redshifts is approximately $\mathcal{R}(z) \sim 5 \times 10^{-2}$ Gyr$^{-1}$. Applying this rate to the quiescent galaxies, we find that correcting for major mergers has only a small impact on the quenched fraction, increasing the number by $\sim$ 10 \%. It is worth noting, however, that while the overall impact is expected to be small, the impact of mergers may not be the same in all environments.

Minor mergers may also occur, and growth due to minor mergers is thought to be similar to that of major mergers \citep[e.g.][]{ownsworth_minor_2014}. However, since the mass bins we use to estimate quenching rates are relatively wide, and minor mergers involve galaxies with mass ratios of 1:10 or higher, they should not cause a significant impact on the results. Galaxies may move to slightly higher masses, but in general this will not have a major effect on the overall number densities of the quiescent populations.      

\subsection{Replenishment of star-forming galaxies}
Another factor to consider is that the pool of star-forming galaxies may not be entirely complete for the redshift intervals studied. While some star-forming galaxies are being quenched and becoming quiescent, others are being formed, or moved to higher stellar mass through star formation, and thus potentially becoming available for quenching in the higher mass bin. We can estimate the upper limit for the number of additional star-forming galaxies that are available to be quenched, using the fact that the star-forming stellar mass function is approximately constant over the redshift range of interest (see Figure \ref{fig:envmfs}). Therefore, to first order, any galaxies quenched are replaced with galaxies of the same stellar mass. To quantify this effect, we can modify Equation \ref{eq:qf} by adding the number of quenched galaxies back into the star-forming sample:

\begin{equation} \label{eq:lambda}
    \lambda_{\textrm{test}} = \frac{\phi_{\textrm{red, low z}}\, -\, \phi_{\textrm{red, high z}}}{[\phi_{\textrm{SF, high z}} \, + (\phi_{\textrm{red, low z}}\, -\, \phi_{\textrm{red, high z}})]\, \Delta t}.
\end{equation}

This reduces our quenched fraction values by $\sim$ 15 percent. However, this correction should represent the maximum amplitude of the effect, since the timescale available for quenching these newly added star-forming galaxies will be shorter than the time interval $\Delta$t.

\subsection{Photometric redshift errors} \label{section:test}
A potential concern with our environmental measures is that photometric redshift errors become larger at high redshift, and may dilute measures of environment relative to those at lower redshift. This could artificially enhance the growth in the passive population at the highest densities when two adjacent redshift bins are compared, since the lower redshift measurements are more precise. To investigate this effect, we broaden the photometric redshifts in a given redshift bin to match the uncertainties at the next highest epoch, e.g. redshifts in the range $0.5 < z < 1$ are decreased in quality (adding additional Gaussian spread) to match the dispersion at $z = 1.25$.
We then redefine our high-, mid- and low-density bins using the $\rho_{250}$ values produced from these new redshift measurements. Running the growth analysis again, we find that redshift errors between epochs produce only a $\sim$ 5 percent increase in our quenched fraction results (Figure \ref{fig:qf}). The trends we see with redshift, mass and density in the original data remain essentially unchanged. The fraction of star-forming galaxies quenched per Gyr remains significantly higher in the densest environments.

\subsection{The effect of including PSBs in the passive sample}
Since PSBs are recently quenched galaxies, we have included them in our quiescent galaxy sample for the analysis of quenching rates. To quantify the impact of leaving PSBs out of our passive sample, and using purely galaxies classified as older quiescent galaxies by the SC method (see Section \ref{section:psbselection}), we repeat our analysis while excluding PSBs. As expected, removing PSBs reduces the measured fraction of quenched galaxies, particularly at high stellar mass and high redshift ($z > 2$), but the influence on the growth rate in the passive population (Figure \ref{fig:gr}) is relatively minor. Consequently, the estimated quenching rate for SF galaxies (Figure \ref{fig:qf}) is also largely unchanged, as is the relative impact of environment. Repeating the analysis outlined in Section 3.4, the ratio of quenching rates in high- to low-density environments changes from 1.7 $\pm$ 0.2 to 1.9 $\pm$ 0.2 for high- mass galaxies (M$_* > 10^{10.7}$ M$_{\odot}$) on average, and from 1.5 $\pm$ 0.1 to 1.6 $\pm$ 0.1 when  galaxies of all masses are considered.

\subsection{The influence of merging halos and cosmic variance}
Overall, our results indicate that the rate of galaxy quenching is significantly higher for galaxies in denser environments. We note, however, that our analysis implicitly assumes that galaxies largely remain within their broad density bins as the Universe evolves, e.g. galaxies in the top third of the density field remain in the top third in the next redshift interval. This simplification ought to be largely robust to the expected impact of galaxy-galaxy mergers (see Section \ref{section:merge}), but may be expected to break down on larger scales, e.g. due to the evolution of large scale structure over cosmic time. Thus one might expect some galaxies in low-density environments to switch to higher density environments at later epochs. Furthermore, our study implicitly assumes that we have surveyed a representative volume of the Universe within each redshift slice, but even within a $\sim$ 1 deg$^2$ field, we can expect some variance due to large-scale structure, particularly for rare massive galaxies \citep[e.g. see][]{moster_cosmic_2011}. Exploring the detailed impact of such effects is beyond the scope of our current analysis, but will be explored using simulated data in future work.

\section{The post-starburst contribution to the growth rate of quiescent galaxies} \label{section:disc}

\begin{figure*}
    \centering
    \includegraphics[width=\textwidth]{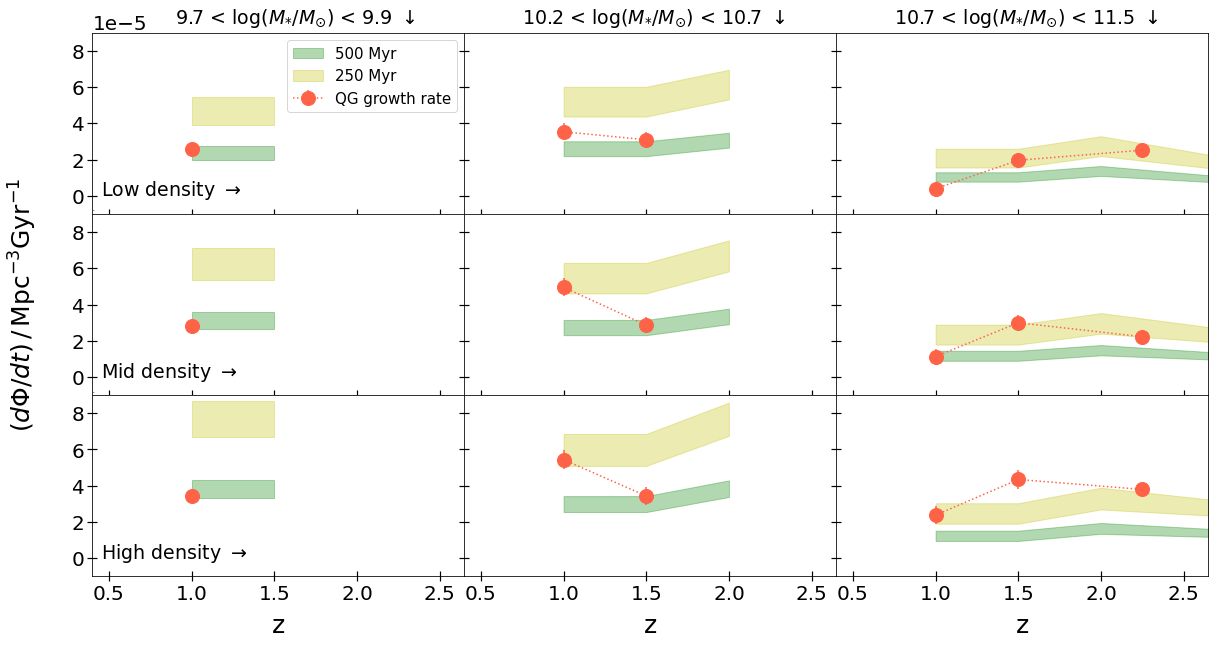}
    \caption{The number density growth rate of quiescent galaxies as a function of redshift (red points), in three bins of mass and environment, compared to the predicted PSB contribution assuming a 250 - 500 Myr lifetime. The yellow band shows the  PSB contribution assuming a 250 Myr lifetime, with the width showing the 1$\sigma$ uncertainty, while the green band show the corresponding contribution assuming a 500 Myr lifetime. Mass and density bins are shown in columns and rows respectively. Errors are propagated from the Poisson error.  }
    \label{fig:psbcont}
\end{figure*}

 Since PSBs have been very recently and rapidly quenched, they prove to be a useful tool for examining the modes of growth for the quiescent population. Following \citet{wild_evolution_2016}, we can compare the growth rate of just quiescent galaxies (Section \ref{section:gr}) to the rate PSBs pass through a given redshift bin to estimate the fraction of quiescent galaxies that pass through a PSB phase. We adopt a typical PSB visibility timescale of $\sim$ 250 -- 500 Myr found by \citet{wild_star_2020} when examining the star formation histories of PSBs, based on fits to spectra and multi-wavelength SEDs. We stress that the visibility timescale is the length of time a galaxy is observed in a PSB phase using the supercolour method, and this timescale may vary in different environments and for different stellar masses depending on burst mass fractions and quenching timescales.
 
 Figure \ref{fig:psbcont} shows the predicted contribution of PSBs to the quiescent galaxy growth rate assuming a 250 - 500 Myr visibility timescale. At high stellar mass  (M$_* > 10^{10.7}$ M$_{\odot}$) we find that  PSBs reproduce roughly half the number density growth rate of quiescent galaxies if the visibility timescale is $\sim$ 500 Myr, or potentially most of the growth with a shorter visibility time. These results are in good agreement with \citet{wild_evolution_2016}, which was based on an earlier version of our UDS catalogue, 
and  
 broadly in agreement with 
 \citet{belli_mosfire_2019}, who found that high-mass PSBs contribute roughly 50\% at $z \sim 2$, falling to roughly 20\% at $z \sim 1.4$.  At intermediate masses we find that PSBs can explain all quiescent galaxy growth, rising to potentially overproducing the red sequence at the lowest stellar masses (M$_* < 10^{10.7}$ M$_{\odot}$) if the visibility timescales are short. 
The potential overproduction at low stellar mass may in part be explained if some of the PSBs arise from rejuvenated systems that were previously quenched  \citep[e.g][]{rowlands_galaxy_2018}.

Interestingly, while the quenching rates are higher in dense environments (see also Figure \ref{fig:qf}), this appears to be matched by the increased contribution from PSBs, at least for the range of stellar masses probed in this analysis
(M$_* > 10^{9.7}$ M$_{\odot}$). Therefore, while quenching is more prevalent in dense environments, the fraction of those galaxies that were 
  {\it rapidly} quenched (according to this simple analysis) does not appear to depend strongly on the density field. At face value this finding appears difficult to reconcile with different quenching mechanisms operating in different environments, unless essentially all quenching mechanisms lead to a PSB phase at these redshifts and stellar masses. It may be that larger samples are required to identify environmental differences in the relative importance of “slow” versus “fast” quenching routes, or perhaps they only become apparent at lower stellar mass. It is notable that the excess of passive galaxies and PSBs in dense environments becomes particularly strong at low stellar mass (Figure \ref{fig:envmfs}), largely below the limit for our evolutionary analysis (M$_* = 10^{9.7}$ M$_{\odot}$).  Future studies with larger and deeper samples may allow a more detailed investigation of the relative importance of different quenching routes in different environments.

\section{Conclusions} \label{section:conc}

In this study, we present an analysis of the build up of the quenched galaxy population in the UKIDSS UDS from $z = 3$ to $z = 0.5$. We use a PCA analysis, first established by \citet{wild_new_2014}, to identify star-forming, quiescent and post-starburst galaxies in the sample. To investigate the influence of environments we use projected galaxy densities to split the catalogue into three environmental bins within broad redshift intervals. The density field and boundaries are defined using galaxies above the same stellar mass at all redshifts, split by projected galaxy density into equal thirds. The key findings are as follows:

\begin{itemize}
    \item A higher fraction of passive galaxies are found in dense environments at all stellar masses in the redshift range $0.5 < z < 2.0$, as seen from a comparison of  stellar mass functions. We also see a sharp up-turn in the relative abundance of low-mass (M$_* < 10^{10}$ M$_{\odot}$) passive galaxies and PSBs in dense environments at $z < 1.5$, which we attribute to the environmental quenching of satellite galaxies entering high-mass halos.
    \item The growth rate in the number density of passive galaxies is steepest in the highest density environments at all masses and redshifts (see Figue \ref{fig:gr}). This difference indicates that star-forming galaxies are quenched more efficiently in these environments.
    \item The growth in the passive population can be used to estimate the fraction of star-forming galaxies quenched per Gyr at different stellar masses, redshifts and environments (see Figure \ref{fig:qf}). We find that the quenching probability increases with both stellar mass and environment. For the most massive galaxies
    (M$_* > 10^{10.7}$ M$_{\odot}$) the quenching probability also increases with redshift, broadly consistent with the quenching probability being proportional to star formation rate.
    \item Averaging over all redshifts and stellar masses, we find that star-forming galaxies are 
    1.5 $\pm$ 0.1 times more likely to quench in the densest third of environments compared to the 
    lowest density third. For the most massive galaxies (M$_* > 10^{10.7}$ M$_{\odot}$) the enhancement factor is 1.7 $\pm$ 0.2.
    Given the dilution effects of projection in our ability to determine environments, these factors are almost certainly lower limits.
    \item 
    At high stellar mass PSBs can account for roughly half of the build-up in the quiescent mass function, assuming a visibility time of 500 Myr, in agreement with previous studies \citep[]{wild_evolution_2016, belli_mosfire_2019}. At low stellar masses the PSB route 
    appears to account for a larger fraction of quiescent galaxies, potentially overproducing the growth rate if the visibility time is significantly shorter than 500 Myr (Figure \ref{fig:psbcont}).
\end{itemize}

This study is a first attempt to separate out the influence of environment on the build-up of the passive and PSB stellar mass functions at high redshift ($z > 1$). We find strong evidence that galaxies quench more rapidly in dense environments at a given stellar mass and redshift. Our study is based on photometric redshifts, however, and therefore quantifying the precise dependence on  environment  is challenging. 
The next generation of multi-object NIR spectrographs (e.g VLT MOONS, Subaru PFS) should produce the next breakthrough in this field, for the first time providing  rest-frame optical spectra for thousands of quenched galaxies at high redshifts. Surveys with these instruments should allow a   more precise characterisation  of the density field, while also allowing a comparison of the detailed astrophysical properties of galaxies in different environments. By comparing star-formation histories, quenching timescales, metallicities, and AGN contributions, it may be possible to separate the detailed influence of environment and 
distinguish between quenching mechanisms.

\section*{Acknowledgements}
We extend our gratitude to the staff at UKIRT for their tireless efforts in ensuring the success of the UDS project. We also wish to recognise and acknowledge the very significant cultural role and reverence that the summit of Mauna Kea has within the indigenous Hawaiian community. We were most fortunate to have the opportunity to conduct observations from this mountain. VW acknowledges STFC grant ST/V000861/1. For the purpose of open access, the authors have applied a creative commons attribution (CC BY) to any journal-accepted manuscript. This work is based in part on observations from ESO telescopes at the Paranal Observatory (programmes 180.A-0776, 094.A-0410, and 194.A-2003). 

\section*{Data Availability}
The imaging data and spectroscopy forming the basis of this work is available from public archives, further details of which can be obtained from the UDS web page (\url{https://www.nottingham.ac.uk/astronomy/UDS/}). A public release of the processed data and photometric redshifts is in preparation. Details can be obtained from Omar Almaini (omar.almaini@nottingham.ac.uk). In the meantime, data will be shared on request to the corresponding author.




\bibliographystyle{mnras}
\bibliography{references} 





\bsp	
\label{lastpage}
\end{document}